\documentclass[aps,prd,twocolumn,superscriptaddress]{revtex4}

\usepackage{amsmath}  
\usepackage{amsfonts}
\usepackage{amssymb}
\usepackage{natbib} 
\usepackage{setspace}
\usepackage{graphicx} 
\usepackage{xspace}
\usepackage{cancel}
\usepackage{url}
\usepackage{color}
\usepackage{float}

\newcommand{\me}{\mathrm{e}} 

\newcommand{\dq}[2]{\hspace{-0.4em}\ensuremath{\frac{\mathrm{d}^{#2}#1}{(2\pi)^{#2}}}\,}
\newcommand{\dx}[1]{\hspace{-0.4em}\ensuremath{\mathrm{d}#1}\,}

\newcommand{\Tr}{\text{Tr}}

\newcommand{\eqn}[1]{Eq.~(\ref{#1})}
\newcommand{\fig}[1]{Fig.~\ref{#1}}
\newcommand{\tab}[1]{Table~\ref{#1}}

\begin{document}

\title{
   A contact interaction model for the $\eta$ and $\eta'$ mesons in a SDE-BSE approach 
   to QCD: masses, decay widths and transition form factors}

\author{Bilgai Almeida Zamora}
\affiliation{Departamento de Investigación en F\'isica, Universidad de Sonora, Boulevard
Luis Encinas J. y Rosales, Colonia Centro, Hermosillo, Sonora 83000, M\'exico}
\author{Enrique Carreon Mart\'inez}
\affiliation{Departamento de F\'isica, Centro de Investigaci\'on y de Estudios Avanzados del
 Instituto Polit\'ecnico Nacional, Apartado Postal 14-740, 07000, Ciudad de M\'exico, M\'exico}
  \author{Jorge Segovia}
\affiliation{Departamento de Sistemas Físicos, Químicos y Naturales, Universidad
Pablo de Olavide, E-41013, Sevilla, Spain.}
\author{J.~J.~Cobos-Mart\'inez\footnote{Corresponding Author}}
\affiliation{Departamento de F\'isica, Universidad de Sonora, Boulevard
Luis Encinas J. y Rosales, Colonia Centro, Hermosillo, Sonora 83000, M\'exico}

\date{\today}

\begin{abstract}
   We construct a contact interaction model for the $\eta$ and $\eta'$ mesons in the SDE-BSE approach to QCD and compute several static properties of these mesons and their transition
   form factors. We find that this model gives an excellent description of the $\eta$ and 
   $\eta'$ static properties, namely their masses, decay width and decay constants. 
   However, a contact interaction disagrees with experimental data for $Q^2$ greater
   than 2 GeV$^2$, and produces transition form factors  in conflict with perturbative QCD
   prediction. This is not surpring and the reasons for this are explained.
\end{abstract}


\maketitle

\date{\today}

\section{Introduction}

Quantum chromodynamics (QCD), the theory of quarks, gluons, and their interactions,
is the accepted theory of the strong interactions at the fundamental level. 
Although QCD dictates the dynamics of quarks and gluons, and therefore, in principle,
the strucure and dynamics of hadrons, which are bound states of quarks and gluons, 
the non-perturbative nature of QCD makes this a difficult problem.
In the low energy, non-perturbative regime of QCD, the emergent phenomena of chiral 
symmetry breaking and confinement govern the static and dynamic properties of hadrons;
for example, these are responsible for 99$\%$ of the mass of the visible universe; see, for 
example, Refs.~\cite{Roberts:2022rxm, Roberts:2023lap} and references therein.

Within the framework of Schwinger-Dyson (SDE) and Bethe-Salpeter equations (BSE) of QCD,
we can investigate the structure and dynamics of hadrons through first principles in the
 continuum. However, drawing a connection between QCD and hadron observables, through 
 the SDE-BSE system, is a difficult task due to the infinite number of equations we need 
 to confront--and  that is  why modelling remains a keystone in hadron physics.
Despite this difficulty, great progress has been made for more than three decades using the 
SDE-BSE approach to non-pertubative QCD and hadron physics, and this has
become a powerful and reliable tool to investigate strong interaction phenomena; see, for 
example, Refs.~\cite{Roberts:1994dr,Maris:2003vk, Bashir:2012fs, Fischer:2018sdj}.
SDE and BSEs for QCD have been extensively applied to the study of hadrons in vacuum and at 
finite density and temperature in a variety of truncation schemes and degrees of 
sophistication; see Ref.~\cite{Fischer:2018sdj} for a recent review.

More than a decade ago,  a simple alternative within the SDE-BSE approach, was proposed,
initially to study pion properties, assuming that quarks interact not via massless vector-boson
 exchange but instead through a symmetry preserving vector-vector contact interaction 
(CI)~\cite{GutierrezGuerrero:2010md,Roberts:2010rn}.  
Over the ensuing years, this contact interaction has been used to investigate static and dynamic
properties of hadrons in vacuum, such as masses, decay constants, electromagnetic elastic and
 transition form factors, parton distribution functions, and generalized parton distribution 
 functions) ~\cite{GutierrezGuerrero:2010md, Roberts:2010rn, Roberts:2011wy, 
 Roberts:2011cf, Chen:2012qr, Chen:2012txa, Segovia:2013uga, Segovia:2014aza, Xu:2015kta,
Bedolla:2015mpa, Bedolla:2016yxq, Raya:2017ggu, Serna:2016kdb, Serna:2017nlr,
Lu:2017cln, Gutierrez-Guerrero:2019uwa, Zhang:2020ecj, Gutierrez-Guerrero:2021rsx,
Xing:2021dwe, Cheng:2022jxe, Wang:2022mrh, Hernandez-Pinto:2023yin, Xing:2022jtt,
Xing:2023eed, Xing:2022mvk} and at finite temperature and density to study, for example, 
the phase diagram of QCD~\cite{Serna:2016ifh, Ahmad:2020jzn,Wang:2013wk}.

The results obtained from the contact interaction model are quantitatively comparable to those obtained using sophisticated QCD model interactions.
Furthermore, the weaknesses and strenghts of this interaction have been identified and
understood.  Despite its weaknesses, the contact interaction has emerged as a powerful tool in the
investigation of hadron properties.  We take this as a sufficient justification to employ this 
interaction in order to construct a contact interaction model for the $\eta$ and $\eta'$ mesons 
in the SDE-BSE approach to QCD and compute several static properties of these mesons and 
their transition form factors. 
We will find  that this model gives an excellent description of the $\eta$ and $\eta'$ static
properties and this makes it an excellent tool for the investigation of $U_A(1)$ symmetry
restoration at finite temperature and density in the SDE-BSE approach to QCD. This will be
reported elsewhere.

This article is organized as follows.  In Sec.~\ref{sec:sde_bse}, we briefly present the 
Schwiger-Dyson and Bethe-Salpeter equations of QCD in rainbow-ladder truncation with 
a contact interaction.
The model parameters for this piece of the interaction are fixed by pion and kaon static
 properties. In Sec.~\ref{sec:ci_rl_nan}, we extend the rainbow-ladder truncation to include a phenomenological
 kernel that represents the non-abelian anomaly. The model parameters for
this part of the interaction are fitted by the masses and decay widths of the $\eta$ and 
$\eta'$ mesons.  
In Sec.~\ref{sec:ci_tff}, we compute the $\eta$ and $\eta'$ transition form
factors using the constructed contact interaction model.
Finally, in Sec.~\ref{sec:summary} we give a summary and conclusions.

\section{\label{sec:sde_bse} SDE-BSE Formalims}

In this section we briefly introduce the SDE-BSE formalism of QCD.  Our focus will be on the
contact interaction since this is employed to produce the results reported in this article.
See Refs.~\cite{Roberts:1994dr,Maris:2003vk, Bashir:2012fs, Fischer:2018sdj} for 
comprenhensive reviews on the SDE-BSE approach to QCD and hadron physics.

 \subsection{\label{sec:sde_ci_rl} The quark SDE and the contact interaction}
 
 The $f$-flavor dressed-quark propagator $S_{f}$ is obtained by solving
  the quark SDE
\begin{eqnarray}
 \label{eqn:quark_sde}
 &&\hspace{-0.5cm} S_{f}^{-1}(p)=i\gamma\cdot p + m_{f} + \Sigma_{f}(p) \,, \\
 \label{eqn:quark_se}
 && \hspace{-0.5cm} \Sigma_{f}(p)=
\int\dq{q}{4}
g^{2}D_{\mu\nu}(p-q)\frac{\lambda^{a}}{2}\gamma_{\mu}S_{f}(q)\Gamma^{a}_{\nu}(p,q)
\,,
\end{eqnarray}
\noindent where $g$ is the strong coupling constant, $D_{\mu\nu}$ is the dressed gluon
 propagator, $\Gamma^{a}_{\nu}$ is the dressed quark-gluon vertex, $m_{f}$  is the $f$-flavor
  current-quark mass and $\lambda^a$ are the usual Gell-Mann matrices. The chiral limit is
  defined by  $m_{f}=0$.

 The SDEs constitute an infinite set of coupled nonlinear integral equations and thus a tractable 
  problem is defined once a truncation scheme has been specified. 
 In practice, this is achieved  by specifying the gluon propagator and the quark-gluon vertex. 
 
  There is extensive literature~\cite{GutierrezGuerrero:2010md, Roberts:2010rn, 
  Roberts:2011wy,  Roberts:2011cf, Chen:2012qr, Chen:2012txa, Segovia:2013uga, 
  Segovia:2014aza, Xu:2015kta, Bedolla:2015mpa, Bedolla:2016yxq, Raya:2017ggu, Serna:2016kdb, Serna:2017nlr, Lu:2017cln, Gutierrez-Guerrero:2019uwa, Zhang:2020ecj, 
  Gutierrez-Guerrero:2021rsx, Xing:2021dwe, Cheng:2022jxe, Wang:2022mrh, 
  Hernandez-Pinto:2023yin, Xing:2022jtt, Xing:2023eed, Xing:2022mvk} where it has been
   shown that  the static properties of low-lying mesons and baryons can be described 
   by assuming that the quarks interact, not via massless vector-boson
    exchange, but instead through a symmetry preserving vector-vector contact interaction
  (CI) with a finite gluon mass
 \begin{eqnarray}
 \label{eqn:contact_interaction}
  g^2D_{\mu\nu}(k) &=& \delta_{\mu\nu} \frac{4\pi
  \alpha_{\rm IR}}{m_g^2} \equiv  \alpha_{\rm eff}\delta_{\mu\nu}
 \,, \\
  \label{eqn:quark_gluon_vertex_rl}
 \Gamma_\mu^a (p,q)&=& \frac{\lambda^a}{2}\gamma_\mu\,,
 \end{eqnarray}
\noindent where $m_g\sim 500$ MeV is an infrared gluon mass scale which is generated
dynamically in QCD~\cite{Bowman:2004jm,Dudal:2008sp,Huber:2010cq,Boucaud:2011ug,
Ayala:2012pb,Gao:2017uox}, and $\alpha_{\rm IR}$ is specified by the strength of the
 infrared interaction in QCD.  There is a critical value of $\alpha_{\rm eff}$ above 
 which chiral symmetry is dynamically broken.

 Equations~(\ref{eqn:contact_interaction}) and (\ref{eqn:quark_gluon_vertex_rl}) specify 
 the so called rainbow truncation of the quark SDE within a contact interaction. 
 This truncation scheme generates a momentum independent dymamical mass $M_f$  
 for the dressed-quark propagator~\cite{GutierrezGuerrero:2010md, Roberts:2010rn}
\begin{equation}
\label{eqn:quark_inverse_ci} S_{f}^{-1}(p)= i \gamma\cdot p +
M_{f} \,.
\end{equation}
\noindent The dynamical, flavor-dependent, constant mass $M_f$ is obtained by solving
\begin{equation}
  \label{eqn:dynamical_mass_ci} 
  M_{f} = m_{f} 
  +  \frac{\alpha_{\rm eff}M_{f}}{3\pi^2}\int_{0}^{\infty}\dx{q^2}\frac{q^2}{q^{2}+M_{f}^{2}}.
\end{equation}
Since the integral in \eqn{eqn:dynamical_mass_ci} is divergent, we must specify a regularization
 prescription.  We use the proper time regularization scheme to write $(q^2+M_f^2)^{-1}$ in
  \eqn{eqn:dynamical_mass_ci} as
\begin{eqnarray}
\frac{1}{q^2+M_f^2}&=&\int^{\infty }_{0} d\tau {\rm e}^{-\tau(q^2+M_f^2)} 
\nonumber \\
&\rightarrow& \int^{\tau_{\text{IR}}}_{\tau_{\text{UV}}} \dx{\tau}\me^{-\tau(q^2+M_f^2)} \nonumber \\
&=&\frac{\me^{-\tau_{\text{UV}}(q^2+M_f^2)} - \me^{-\tau_{\text{IR}}(q^2+M_f^2)}}{q^2+M_f^2}.
\label{eqn:ptr} 
\end{eqnarray}
\noindent In \eqn{eqn:ptr}, $\tau_{\text{IR}}^{-1}\equiv \Lambda_{\text{IR}}^2$ and
$\tau_{\text{UV}}^{-1} \equiv \Lambda_{\text{UV}}^2$  are infra-red and ultra-violet regulators,
 respectively, which will be specified later. A non-zero value for $\tau_{\text{IR}}$ implements
  confinement by ensuring the absence of quarks production thresholds~\cite{Ebert:1996vx}.  
It has been shown that an excitation described by a pole-less propagator would never reach its
 mass-shell~\cite{Ebert:1996vx}. Moreover, since  the CI is not renormalizable, 
$\Lambda_{\text{UV}}$ cannot be removed, but instead plays a dynamical role and sets the 
scale for all dimensioned physical uantities. 
After integration over $q^2$,  \eqn{eqn:dynamical_mass_ci} can be written as
\begin{equation}
  \label{eqn:dynamical_mass_ci_ptr}
 M_{f}= m_{f} + \frac{\alpha_{\rm eff}M_f}{3\pi^2}
 \int_{\tau_{\text{UV}}}^{\tau_{\text{IR}}}\dx{\tau}\tau^{-2}\me^{-\tau M_f^2}.
\end{equation}
By solving \eqn{eqn:dynamical_mass_ci_ptr} we can obtain the dynamical mass of the
quark of flavor $f$.

\subsection{\label{sec:bse-ci_rl} The BSE and the contact interaction}

In quantum field theory,  meson bound states in a given $J^{PC}$ channel,  whose flavor
 structure  is given by  a non-perturbative dressed quark-antiquark pair ($f\bar{g}$),  
 are described by the  Bethe-Salpeter equation (BSE)~\cite{Nambu:1950dpa,Salpeter:1951sz, Gell-Mann:1951ooy, Nakanishi:1969ph}
\begin{equation}
\label{eqn:bse} 
\left[\Gamma_{H}(p;P)\right]_{tu}=
\int\dq{q}{4}\left[K(p,q;P)\right]_{tu}^{rs}\left[\chi(q;P)\right]_{sr} \,,
\end{equation}
\noindent where $\chi(q;P)=S_{f}(q_{+})\Gamma_{H}(q;P)S_{g}(q_{-})$; $q_{+}=q+\eta
P$, $q_{-}=q-(1-\eta)P$; $\eta \in [0,1]$ is a momentum-sharing parameter, $p$ ($P$) is the relative
 (total) momentum of the quark-antiquark system, with $P^{2}=-m_{H}^{2}$ and $m_{H}$ the mass
 of the meson ; $S_{f (g)}$ is the non-perturbative $f (g)$-flavor
  dressed-quark propagator, already discussed; $\Gamma_{H}(p;P)$ is the meson Bethe-Salpeter
   amplitude (BSA),  where $H$ specifies the quantum numbers and flavor content of the 
   meson; $r,s,t$, and $u$ represent color, flavor, and spinor indices; and $K(p,q;P)$ 
   is the quark-antiquark scattering kernel.

 Equations~(\ref{eqn:contact_interaction}) and (\ref{eqn:quark_gluon_vertex_rl}) define the kernel of the
 quark SDE. But they also define the kernel of the BSE, \eqn{eqn:bse}, through the axial-vector 
 Ward-Takahashi identity (axWTI)~\cite{Maris:1997hd}
\begin{equation}
  \label{eqn:axwti}
  - i P_{\mu}\Gamma_{5\mu}(k;P)=S^{-1}(k_{+})\gamma_{5} + \gamma_{5}S^{-1}(k_{-}).
\end{equation}
\noindent This identity, which encodes the phenomenological features of dynamical chiral symmetry
 breaking in QCD, relates the axial-vector vertex, $\Gamma_{5\mu}(k;P)$, to the quark propagator, 
 $S(k)$, which in turn  implies a relationship between the kernel in the BSE, \eqn{eqn:bse}, and that in 
 the quark SDE, \eqn{eqn:quark_sde}. This relation must be preserved by any viable truncation scheme 
 of the SDE-BSE coupled system, thus constraining the content of the quark-antiquark scattering kernel
  $K(p,q;P)$. For the CI, \eqn{eqn:axwti} implies that the quark-antiquark scattering kernel $K(p,q;P)$
  is given by
\begin{equation}
  \label{eqn:bse_kernel_rl_ci} \left[K_{\rm L}(p,q;P)\right]_{tu}^{rs}= -g^{2}D_{\mu\nu}(p-q)
  \left[\frac{\lambda^{a}}{2}\gamma_{\mu}\right]_{ts}
  \left[\frac{\lambda^{a}}{2}\gamma_{\nu}\right]_{ru} \,,
\end{equation}
\noindent where $g^{2}D_{\mu\nu}(k)$ is given by \eqn{eqn:contact_interaction}.

Thus, the homogeneous BSE (we use $\eta=1$ in this work), in rainbow-ladder (RL) truncation, with a contact interaction, takes the simple form
\begin{equation}
  \label{eqn:bse_rl_ci}
  \Gamma_{H}(p;P)=-\frac{4}{3}\alpha_{\rm eff}\int\dq{q}{4}
  \gamma_{\mu}S_{f}(q_{+})\Gamma_{H}(q;P)S_{g}(q_{-})\gamma_{\mu}.
\end{equation}
We note that some of the integrals that appear in \eqn{eqn:bse_rl_ci} are also divergent but 
will  be regulated using the propertime regularization scheme, \eqn{eqn:ptr}, as  
we will do with all divergent integrals in this work.

Since the kernel of \eqn{eqn:bse_kernel_rl_ci} in  RL truncation with a contact interaction  
does not depend on the external relative momentum, a symmetry-preserving regularization 
will give solutions which are also independent of it. Therefore, for example, the general form of 
the BSA for a  pseudoscalar meson is given by
\begin{equation}
  \label{eqn:ps_bsa_rl_ci}
  \Gamma_{\text{Ps}}(P)= \gamma_{5}\left[ i E_{\text{Ps}}(P) + \frac{1}{2M_{\text{R}}}\gamma\cdot P F_{\text{Ps}}(P)\right],
\end{equation}
\noindent where $M_{\text{R}}= M_f M_g/(M_f+M_g)$. We are interested only in pseudoscalar
mesons but similar expressions to \eqn{eqn:ps_bsa_rl_ci} can be written down for other channels.

The BSE is a homogeneous equation and thus the BSA has to be normalized by a separate
 condition.   In the RL approximation, the normalization  condition is ~\cite{Nambu:1950dpa,Salpeter:1951sz, Gell-Mann:1951ooy, Nakanishi:1969ph}:
\begin{equation}
  \label{eqn:RLNorm}
  1=N_{c}\frac{\partial}{\partial P^{2}}\int\dq{q}{4}
  \text{Tr}\left[\overline{\Gamma}_{H}(-Q)S(q_{+})\Gamma_{H}(Q)S(q_{-})\right]
  |_{Q=P} \,,
\end{equation}
\noindent where $P^{2}=-m_{H}^{2}$, $\Gamma_{H}$ is the normalized BSA of the meson $H$, 
and  $\overline{\Gamma}_{H}$ is its charge-conjugated version. 

Once the BSA has been canonically-normalized, we can compute observables with it. For example,
the leptonic decay constant of  a pseudoscalar meson, $f_{\text{Ps}}$,  can be calculated 
from
\begin{equation}
  \label{eqn:ps_decay_constant}
 P_{\mu}f_{\text{Ps}}= N_{c}\int\dq{q}{4}\text{Tr}\left[\gamma_{5}\gamma_{\mu}S_{f}(q_{+})
 \Gamma_{\text{Ps}}(P)S_g(q_{-})\right], 
\end{equation}
\noindent where the trace is over Dirac indices. 

\subsection{\label{sec:rl_ci_parameters} Numerical results with the RL contact interaction}

We work in the isospin symmetric limit, where $m_{u}=m_d$ and use the notation 
$m_l\equiv m_u$ for the current mass of the light quarks. The model parameters
in the RL truncation are thus
 $m_l$, the strange current quark mass $m_s$, the effective coupling $\alpha_{\rm eff}$
  (or $\alpha_{\text{IR}}$ since $m_g$ is fixed to 500 MeV), and the ultraviolet regulator 
  $\Lambda_{\rm UV}$. The infrared regulator $\Lambda_{\rm IR}$ is fixed to approximatelly
  $\Lambda_{\rm QCD}= 240 $ GeV. 
 The three parameters $m_l$, $\alpha_{\rm eff}$ (or $\alpha_{\text{IR}}$ ), and 
 $\Lambda_{\rm UV}$ were fixed in Ref.~\cite{GutierrezGuerrero:2010md} from the pion
  mass,  pion decay constant,  and chiral condensate using a least-squares procedure. 
  We use a normalization of the BSA amplitude such that the experimental pion decay constant
  is $f_\pi= 93$ MeV. 
  Ref.~\cite{GutierrezGuerrero:2010md} gives $m_l= 8 $ MeV, 
  $\alpha_{\rm eff}= 1/(110\, \rm MeV)^2 = 8.3 \times 10^{-5}$ MeV$^{-2}$ (this gives
  $\alpha_{\text{IR}}=0.52\pi$ for $m_g=500$ MeV), and $\Lambda_{\rm UV}= 823$ MeV.
  
 In order to determine the current mass of the strange quark, we fit experimental value of kaon 
 mass, $m_K=497$ MeV. This gives $m_s= 187$ MeV, and we predict the value $f_K=96$ MeV 
 for  the kaon decay constant. The value obtained here for $f_K$ is similar to the one obtained in
 the NJL model of QCD, where it is found that $f_K=91$ MeV~\cite{Hutauruk:2019ipp}.
 Recent experimental analyses of the current quark masses and pseudoscalar meson leptonic 
 decay constants have found  $m_s/m_l$= $27.33_{-0.77}^{+0.67}$ and 
 $f_K/f_\pi=1.193(2)$~\cite{Workman:2022ynf}. Our results for these ratios are
 $m_s/m_l=23.375$ and $f_K/f_\pi=1.03$, which reasonably agree with the experimental values.
  In \tab{tab:rl_ci_masses} we give a summary of the values for the CI-RL model parameters.  
\begin{table}[H]
\centering
\begin{tabular}{ccccc}
\hline \hline
      & $m_{\pi}$ & $m_{K}$ & $f_{\pi}$ & $f_{K}$ \\ 
      \hline
 PDG~\cite{Workman:2022ynf} & 140   & 497  & 93     & 110   \\ 
 \hline
 RL-CI & 141   & 500  & 94     & 96   \\
 \hline \hline
\end{tabular}
\caption{Results for pion and kaon static properties (in MeV) obtained with  $m_l= 8$ MeV, 
$m_s=187$ MeV,  $\alpha_{\rm eff}= 1/(110 \rm MeV)^2 = 0.91\times 10^{-4}$ MeV$^{-2}$, 
and $\Lambda_{\rm UV}= 823$ MeV. The dynamical masses for the light and strange quarks are
$M_l=410$ MeV, $M_s=557$ MeV, respectively.  The values in the second row are taken 
from the  Particle  Data Group~\cite{Workman:2022ynf}.}
\label{tab:rl_ci_masses}
\end{table}    
\section{\label{sec:ci_rl_nan} Including the Non-Abelian Anomaly}

The RL kernel is insufficient to describe the $\eta$ and $\eta'$ properties, since, 
for example,  it does not produce mixing between  $u\bar{u}$, $d\bar{d}$ and $s\bar{s}$ correlations.  A way to introduce  mixing between these correlations, and therefore improve the description of the $\eta$ and $\eta'$ mesons, is to go beyond the RL truncation and add to $K_L$, given in \eqn{eqn:bse_kernel_rl_ci}, a term that represents the non-Abelian Anomaly; that is
\begin{equation}
\label{eqn:bse_kernel}
K(p,q;P)= K_{\rm L}(p,q;P) + K_{A}(p,q;P)
\end{equation}
where~\cite{Ding:2018xwy}
 \begin{eqnarray}
  \label{eqn:bse_kernel_A} 
  \left[K_{\rm A}(p,q;P)\right]_{tu}^{rs}&=& \xi
  \left(\cos^2\theta\left[\mathcal{Z}i\gamma_{5}\right]_{rs}\left[\mathcal{Z}i\gamma_{5}\right]_{tu} \right. \nonumber \\
  &+& \sin^2\theta \left.\left[ \hat{\mathcal{Z}}\cancel{P}\gamma_{5}\right]_{rs}
  \left[\hat{\mathcal{Z}}\cancel{P}\gamma_{5}\right]_{tu} \right)
 \end{eqnarray}
 with $\mathcal{Z}={\rm diag}(1,1,M_l/M_s)$ and $\hat{\mathcal{Z}}=(1/M_l)\mathcal{Z}$,
 matrices in flavor space.
 Here, the dynamical masses of the light, $M_l$, and strange, $M_s$, quarks
 are fixed only from pion and kaon phenomenology. We note that the origin of $K_A$ is 
 phenomenological; see Ref.~\cite{Ding:2018xwy}
 
 In \eqn{eqn:bse_kernel_A}, the model parameters are $\xi$ and $\theta$: $\xi$
 is a dimensionless coupling strength and $\theta$ controls the relative strength of 
 the two tensor structures. In principle, $\xi$ would also depend on the relative momenta
between the quark and the antiquark; however, to be consistent with the contact
interaction RL kernel, we requiere $\xi$ to be a constant.  Then, this gives a BSA for the
$\eta$ and $\eta'$ independent of the relative momentum between the quark and the
antiquark, which is a signature of a contact interaction. 
The model parameters  $\xi$ and $\theta$ are determined from experimental values of the 
masses, decay constants, and decay widths of the $\eta$ and $\eta'$.
     
 \subsection{\label{sec:formulas} Leptonic decay constants and  2-photon widths for the 
 $\eta$ and  $\eta'$}
 
   We now discuss the $\eta$ and $\eta'$ static properties in our contact interaction model. 
   For this, it is  convenient to  work with  the $N_f=3$ quark flavor basis, where the $\eta$ 
   and $\eta'$ wave functions can be written as~\cite{Feldmann:1998sh,Feldmann:1998vh,
   Feldmann:1999uf}
  \begin{equation}
  \label{eqn:eta-etaprime-wfs}
  \chi_{h}(p;P)=d_l\,\chi^{l}_{h}(p;P) + d_s\,\chi^{s}_{h}(p;P), \hspace{3mm}
  h= \eta,\,\eta'\,
  \end{equation}
  where $d_l={\rm diag}(1,1,0)$,  $d_s= {\rm diag}(0,0,\sqrt{2})$ are matrices in flavor space.
  Here $\chi^{l}_{h}(p;P)$ and $\chi^{s}_{h}(p;P)$ are Bethe-Salpeter wave functions for 
  the $l\overline{l}$ and $s\overline{s}$ correlations in the $\eta$ and $\eta'$ mesons,
  and the correspoding Bethe-Salpeter amplitudues are obtained from
   \begin{equation}
   \label{eqn:eta-etaprime-wfs-flavor}
    \chi_{h}^{f}(p;P)= S_{f}(p_{+})\Gamma^{f}_{h}(p;P)S_f(p_{-}),
   \end{equation}
  where $f=l,\,s$. Thus, the BSA of the $\eta$ and $\eta'$ are given by ($h= \eta,\,\eta'$)
  \begin{equation}
  \label{eqn:BSA_h}
  \Gamma_{h}(p;P)=d_l\,\Gamma^{l}_{h}(p;P) + d_s\,\Gamma^{s}_{h}(p;P).
  \end{equation}
  Inserting equations~(\ref{eqn:eta-etaprime-wfs}), ~(\ref{eqn:eta-etaprime-wfs-flavor}) and ~(\ref{eqn:BSA_h}) into the Bethe-Salpeter equation, \eqn{eqn:bse},  we obtain a set of coupled equations for the light and strange correlations $\Gamma^{l}_{h}$ and  $\Gamma^{s}_{h}$ in 
  the $\eta$  and $\eta'$ mesons, which can be solved by matrix methods~\cite{Bedolla:2015mpa}:
 \begin{align}
 \label{eqn:bse_l}
    \Gamma_{h}^l(p;P) =& - \frac{4}{3} \int \frac{\mathrm{d}^4q}{(2\pi)^4} \; g^2D_{\mu \nu} (k-q) \gamma_{\mu} \chi^{l}_{h} (q;P) \gamma_{\nu} \nonumber \\
    &+ \xi\int \frac{\mathrm{d}^4q}{(2\pi)^4} \; \cos^2 \theta  \Tr \left[ Z \gamma_5 \chi_{h}(q;P)\right]i\gamma_5 \nonumber \\ 
    & + \xi \int \frac{\mathrm{d}^4q}{(2\pi)^4} \;  \frac{\sin^2 \theta}{M_l^2} 
    \Tr \left[ Z \gamma_5 \cancel{P} \chi_{h}(q;P)\right]\gamma_5 \cancel{P} 
    \\
\label{eqn:bse_s}
    \Gamma_{h}^s(p;P) =& - \frac{4}{3} \int \frac{\mathrm{d}^4q}{(2\pi)^4} \; 
    g^2D_{\mu \nu} (k-q) \gamma_{\mu} \chi^{s}_{h} (q;P) \gamma_{\nu} \nonumber \\
    & +  \frac{\xi \nu_A}{\sqrt{2}} \int \frac{\mathrm{d}^4q}{(2\pi)^4} \; \cos^2 \theta 
    \Tr \left[ Z \gamma_5 \chi_{h}(q;P)\right] i\gamma_5 \nonumber \\     
    & + \frac{\xi \nu_A}{\sqrt{2}}  \int \frac{\mathrm{d}^4q}{(2\pi)^4} \; 
    \frac{\sin^2 \theta}{M_l^2} \Tr \left[ Z \gamma_5 \cancel{P} \chi_{h}(q;P)\right]
     \gamma_5 \cancel{P},
\end{align}
where the trace is over flavor and Dirac indices. Indeed, these these are a set of coupled 
equations for  $\Gamma^{l}_{h}$ and  $\Gamma^{s}_{h}$ since $\chi_h(q;P)$ ($h=\eta,\eta'$), 
 given by \eqn{eqn:eta-etaprime-wfs-flavor}, contains both correlations and thus 
produces mixing.  Since the non-Abelian kernel does not introduce any dependence in the
 relative momenta between the quark and the antiquark (recall that we take $\xi$ to be
  independent of momenta), the BSA for the correlations $\Gamma^{l}_{h}$ and 
  $\Gamma^{s}_{h}$ has the general structure:
\begin{equation}
 \label{eqn:ps_bsa_rl_A_ci}
  \Gamma_{h}^{f}(P)= \gamma_{5}\left[ i E_{h}^{f}(P) 
  + \frac{1}{M_f}\cancel{P}F_{h}^{f}(P)\right]
\end{equation}
for $h=\eta,\eta'$ and $f=l,\,s$. 

Using standar projection methods~\cite{Bedolla:2015mpa}, equations~(\ref{eqn:bse_l}) 
and~(\ref{eqn:bse_s}) can be written terms of the pseudoscalar ($E_h^{l,\,s}$) and pseudovector
($F_h^{l,\,s}$) components, which in turn can be written as a eigenvalue equation for the vector 
$\left(E_h^l,\; F_h^l,\; E_h^s,\; F_h^s\;  \right)^T$ for $h=\eta,\eta'$:
%
%
\begin{equation}
\label{eqn:mat_eqn}
\begin{pmatrix}
E_h^l \\
F_h^l \\
E_h^s \\
F_h^s
\end{pmatrix}
= 
\left[K_h(P^2)\right]_{4\times 4}
\begin{pmatrix}
E_h^l \\
F_h^l \\
E_h^s \\
F_h^s
\end{pmatrix}
\end{equation}
where the matrix elements of $K_h(P^2)$ will be reported elsewhere.
We note that the mixing between the light and strange correlations is proportional to $\xi$. This mixing vanishes when $\xi=0$ and the light and strange correlations decouple, since $K_A=0$ and $K_h$ reduces to the RL kernel, $K_{h}=K_L$. 

The masses of the $\eta$ and $\eta'$ are computed from the condition
\begin{equation}
\label{eqn:det_eqn}
\text{det}\left[K_h(P^2=m_h^2)-I_{4\times 4}\right]=0.
\end{equation}
That is, we adjust $m_h$, for $h=\eta,\,\eta'$, until \eqn{eqn:det_eqn} is satisfied.

Since the kernel now depends on the meson momentum $P$, the canonical normalization
condition is more complicated. For this reason, it is useful to introduce an
alternative, but equivalent, normalization condition. To this end, and eigenvalue
$\lambda(P^2)$ is introduced on the left-hand side of \eqn{eqn:mat_eqn} such that
it has solutions for all $P$. In terms of  $\lambda(P^2)$ the bound state condition now becomes
\begin{equation}
\label{eqn:eig_eqn}
\lambda(P^2=m_h^2)=1.
\end{equation}
That is, we adjust $m_h$, for $h=\eta,\,\eta'$, until \eqn{eqn:eig_eqn} is satisfied.
The smallest mass is identified with the $\eta$ mass and the largest with that of the
$\eta'$.

In terms of $\lambda(P^2)$, the normalization condition for the BSA 
is~\cite{Nakanishi:1965zza,Nakanishi:1965zz}

%
%
\begin{align}
  \label{eqn:alt_norm_cond}
  \left[\frac{\mathrm{d \ln}\,\lambda(P^{2})}{\mathrm{d}P^{2}}\right]^{-1}_{P^{2}= m_h^{2}} = 2\Tr \int & \frac{\mathrm{d}^{4}q}{(2\pi)^{4}} 
  \left[ \Bar{\Gamma}^{l}_{h} (-P) \chi^{l}_{h} (q;P) \right. \nonumber \\ 
    & + \left.\Bar{\Gamma}^{s}_{h} (-P) \chi^{s}_{h} (q;P) \right],
\end{align}
where trace is over color and Dirac indices.

Once the BSA has been normalized, observables such as the lepton decay constants 
and 2-photon decay widths can be calculated.  The decay widths $h \to \gamma \gamma$  can
be computed from~\cite{Ding:2018xwy}
\begin{equation}
    \Gamma_{h\to\gamma\gamma} = \frac{9\alpha_{\text{em}}^2}{64\pi^3} m^3_h 
    \left[ c_l \frac{f^l_h}{(f^l)^2} + c_s \frac{f^s_h}{(f^s)^2} \right]^2,
\end{equation}
with $\alpha_{em}=1/137$, $c_l=5/9$, $c_s=\sqrt{2}/9$ and 
$(f^l)^2 =  (f^l_{\eta})^2 + (f^l_{\eta'})^2$, $(f^s)^2 =  (f^s_{\eta})^2 + (f^s_{\eta'})^2$
%
%
The leptonic decay constants $f_h^l$ and $f_h^s$, for $h=\eta,\,\eta'$, for the light
and strange correlations $\Gamma_h^l$ and $\Gamma_h^s$, respectively, can be
obtained from \eqn{eqn:ps_decay_constant} with $g=f=l,\,s$, and 
 $\Gamma_h^f(P)=\Gamma_{\text{Ps}}(P)$ ($f=l,\,s$).

\subsection{\label{sec:resultst_!} Numerical results for the $\eta$ and $\eta'$ masses, leptonic decay constants, and 2-photon decay widths}

The model parameters for the Rainbow-Ladder part of the kernel, together with the
current masses of the light and strange quarks, are fixed by
pion and kaon static properties; see \tab{tab:rl_ci_masses}. The remaining parameters
to be fixed are $\xi$ and $\theta$ and these define the non-Abelian contribution to the
 kernel. We fix these parameters from experimental values of  the masses, decay constants, 
 and decay widths of the $\eta$ and $\eta'$, using a least-squares procedure.
 The experimental  values of these constants are given 
 in \tab{tab:expt_eta_etaprime_1}.
 \begin{table}[H]
\centering
\begin{tabular}{ccccc}
\hline \hline
 & $m_\eta$ & $m_{\eta'}$ & $\Gamma_{\eta\gamma\gamma}$ & 
 $\Gamma_{\eta'\gamma\gamma}$ \\ \hline
PDG~\cite{Workman:2022ynf}. & 548  & 958  & 0.516(22)  & 4.35(36) \\ 
 \hline \hline
\end{tabular}
\caption{Experimental values for the masses, decay widths, and leptonic decay constants for $\eta$ and $\eta'$ mesons. All quantities are in MeV, except the decay widths which are in keV.}
\label{tab:expt_eta_etaprime_1}
\end{table}

We now fix $\xi$ and $\theta$ by minimising the root-mean-square 
fractional error for the $\eta$ and $\eta'$ masses and decay widths. The results for
the parameters found in this way are given in the second row of \tab{tab:K_A_Pars1},
and the corresponding values for the $\eta$ and $\eta'$ masses and decay widths
are given in the first row of \tab{tab:eta-etaprime_results}. As can be seen, the
results are in excellent agreement with experimental data, except for the decay width of
the $\eta$ meson, but we can do better as we now explain.
\begin{table}[h]
\centering
\begin{tabular}{c|ccc}
\hline \hline
& $\xi$ & $\cos^2 \theta$ & $\Lambda_{UV}$ \\ 
 \hline
CI model (fit-I)   & 8.15  & 0.898  & 823  \\ 
\hline
CI model (fit-II)  & 5.54  & 0.999  & 810  \\ 
\hline \hline
\end{tabular}
\caption{\label{tab:K_A_Pars1} Parametesr for the non-Abelian anomaly contribution
 the kernel. $\Lambda_{\text{UV}}$ is given in MeV.}
\end{table}
\begin{table}[h]
\centering
\begin{tabular}{lllll}
\hline \hline
& $m_\eta$ & $m_{\eta'}$ & $\Gamma_{\eta\gamma\gamma}$ & $\Gamma_{\eta'\gamma\gamma}$ \\ 
\hline
PDG \cite{Workman:2022ynf} & 548  & 958  & 0.516(22)   & 4.35(36) \\
CI model (fit-I)  & 548  & 920  & 0.287  & 4.62  \\
CI model (fit-II) & 558  & 920  & 0.418  & 4.16   \\
Ding \cite{Ding:2018xwy}  & 560  & 960  & 0.420   & 4.66 \\
Osipov \cite{Osipov:2014dya}  & 547  & 930  & 0.523  & 5.22 \\
Takizawa \cite{Takizawa:1995ku}  & 510    & -- & 0.503$^*$  & -- \\ 
\hline \hline
\end{tabular}
\caption{\label{tab:eta-etaprime_results} Masses and decay widths for $\eta$ and 
$\eta'$ mesons. All quantities are in GeV, except the decay widths are in keV.}
\end{table}

The UV cutoff parameter, $\Lambda_{\text{UV}}$, was introduced in \eqn{eqn:ptr} to regularize
the divergent integrals appearing in the quark SDE and meson BSE.  
Together with $\alpha_{\text{IR}}$, these parameters specify the RL contact
 interaction model. Recall that these parameters were determined using pion and kaon static
  properties. Similar, divergent integrals to \eqn{eqn:dynamical_mass_ci} also appear in
the non-Abelian part of the kernel, see equations~(\ref{eqn:bse_l}) and~(\ref{eqn:bse_s}).
Thus, in order to have a better description of the masses {\it and} decay widths
for the $\eta$ and $\eta'$ mesons, we introduce another ultraviolet cutoff 
$\widetilde{\Lambda}_{\text{UV}}$ into the divergent integrals that appear in the 
non-Abelian part of the kernel in equations~(\ref{eqn:bse_l}) and~(\ref{eqn:bse_s}),
and fix $\xi$, $\theta$, and $\widetilde{\Lambda}_{\text{UV}}$ by minimising the 
root-mean-square fractional error for the $\eta$ and $\eta'$ masses and decay widths.
This does not affect the results obtained for the pion and kaon.
The reason for doing this is that the decay constants, and thus the decay widths,  are sensitive to 
the ultraviolet cutoff  used in the proper time regularization 
The results for the parameters, found in this way, are given in the second row of
 \tab{tab:K_A_Pars1}, and the corresponding values for the $\eta$ and $\eta'$ masses
  and decay widths are given in the third row of \tab{tab:eta-etaprime_results}. 
  Clearly, our results agree nicely with experimental data for all
  four observables. In \tab{tab:eta-etaprime_results}, we also list the results obtained
  in other approaches. The corresponding leptonic
  decay constants are given in \tab{tab:eta-etaprime_decay_const} and also are  
  really good agreement with experimental data.
  
With all the parameters fixed, we now proceed to compute the transition form factors
for the $\eta$ and $\eta'$ mesons.
 \begin{table}[h]
\centering
\begin{tabular}{ccccc}
\hline \hline
& $f_{\eta}^l$ & $f_{\eta}^s$ & $f_{\eta'}^l$ & $f_{\eta'}^s$ \\ 
\hline
PDG \cite{Workman:2022ynf}  & 0.090(13)    & -0.093(28)   & 0.073(14)  & 0.094(8) \\
CI model (fit-I)  & 0.074   & -0.060  & 0.078  & 0.086  \\
CI model (fit-II)  & 0.081   & -0.049   & 0.073   & 0.097   \\
Ding \cite{Ding:2018xwy}    & 0.074   & -0.094   & 0.068  & 0.101   \\ 
\hline \hline
\end{tabular}
\caption{\label{tab:eta-etaprime_decay_const} Leptonic decay constants for the $\eta$ 
and $\eta'$ mesons. All quantities are in GeV.}
\end{table}
\section{\label{sec:ci_tff} Transition form factors for the $\eta$ and $\eta'$ mesons
in a contact interaction}

The transition process $\gamma^{*}\gamma\to h$ ($h=\eta,\,\eta'$) is characterised 
by a single transition form factor $G_h(Q^2)$, where $Q^2$ is the photon virtuallity. 
For each meson, the transition form factor can be computed from, in a rainbow-ladder
truncation, by~\cite{Ding:2018xwy}
\begin{align}
\label{eqn:TFF_def}
    & \frac{\alpha_{\text{em}}}{2\pi} \epsilon_{\mu \nu \alpha \beta}q_{1\alpha}q_{2\beta}
    G_h (q_{1}^2,q_1\cdot q_2,q_2^2)  \nonumber \\ 
  & =  \frac{\alpha_{\text{em}}}{2\pi}\epsilon_{\mu \nu \alpha \beta}q_{1\alpha}q_{2\beta} 
  \left[G_h^l(Q^2)  + G_h^s(Q^2)  \right] \nonumber \\
  &  = \Tr_{\text{D}} \int \frac{\mathrm{d}^4q}{(2\pi)^4} \left[i c_l\chi^l_\mu(k,k_1) \Gamma_h^l(k_1,k_2)S_l(k_2) i\Gamma_\nu^l(k_2,k)\right. \nonumber \\
   & \left. + \;i c_s\chi^s_\mu(k,k_1) \Gamma_M^s(k_1,k_2)S_s(k_2) i\Gamma_\nu^s(k_2,k)\right]
\end{align}
where the trace is over Dirac indices; the momentum distribution is $k_1=k+q_1$ 
and $k_2=q-q_2$; and the kinematic conditions are $q_1^2=Q^2$, $q_2^2=0$, and 
$2q_1\cdot q_2=-(Q^2+m_h^2)$ ($h=\eta,\,\eta'$).
$\chi^f_\mu(k,p)=S_f(k)i\Gamma_{\nu}^f(k,p)S_f(p)$ ($f=l,\,s$) is  the unamputated quark-photon vertex, and $\Gamma_\nu^f$ the  (amputated) quark-photon vertex. In \eqn{eqn:TFF_def}, all
quantities have been determined earlier, namely we have already determined the model parameters
and the masses of the $\eta$ and $\eta'$ mesons. Furthermore, we have  obtained previously 
all quark propagators and Bethe-Salpeter amplitudes by solving their respective integral equations.
Altough the quark-photon vertex has been obtained in previous work, see for example,
Ref.~\cite{Bedolla:2016yxq}, we discuss it briefly for completeness.

\subsection{\label{sec:quark-photon-vertex} The quark-photon vertex}

The coupling of a photon with the bound state’s charged constituent is given by 
the quark-photon vertex $\Gamma_\nu^f(p,k; Q)$, where $f$ denotes the flavor of the
quark that interacts with the photon, $p$ ($k$) is the incoming (outgoing) quark momentum and $Q=p-k$ is the photon momentum. The quark-photon vertex satisfies 
its own SDE but it is also constrained by the gauge invariance of quantum electrodynamics
 through the Ward-Takahashi identity
\begin{equation}
\label{eqn:wti}
iQ_\mu\Gamma_\mu(k,p;Q)= S^{-1}(k)-S^{-1}(p).
\end{equation}
Satisfying this identity, and its $Q\to 0$ limit, is crucial for the conservation of the
electromagnetic current ~\cite{Bedolla:2016yxq}.

In the RL truncation with a contact interaction, the SDE for the quark-photon vertex
is~\cite{Bedolla:2016yxq}
\begin{equation}
\label{eqn:qpv_sde}
\Gamma_\mu(Q)= \gamma_\mu -\dfrac{4}{3}\alpha_{\text{eff}}
\int \frac{\mathrm{d}^4q}{(2\pi)^4} 
\gamma_\nu S(q + Q)\Gamma_{\mu}(Q)S(q)\gamma_\nu.
\end{equation}
Note that the RL truncation together with the contact interaction gives a quark-photon
vertex that is independent of the relative momenta between the quark and the 
antiquark, and thus $\Gamma_\mu$ depends only on the 
photon momenta $Q$~\cite{Bedolla:2016yxq}. Therefore, the quark-photon vertex is given by
\begin{equation}
\label{eqn:qpv}
\Gamma_\mu(Q)=\gamma_\mu^T P_T(Q^2) + \gamma_\mu^L P_L(Q^2),
\end{equation}
where $Q_\mu\Gamma_\mu^T=0$, and $\gamma_\mu^T + \gamma_\mu^L
=\gamma_\mu$.
The functions $P_T(Q^2)$ and $P_L(Q^2)$ can be found from \eqn{eqn:qpv_sde} using
standard projection methods; see Ref~\cite{Bedolla:2016yxq} for details. It is found that 
$P_L=1$ and 
\begin{equation}
P_T(Q^2)=\dfrac{1}{1-K_V(Q^2)}
\end{equation}
where $K_V(Q^2)$ is the Bethe-Salpeter bound state kernel in the vector channel within the 
RL truncation of the SDE-BSE; see equations (A10)-(A13) in Ref.~\cite{Bedolla:2015mpa}. 
Thus, because of the dressing of the quark-photon vertex, electromagnetic elastic and transition
 form factors will have a pole at  $Q^2=-m_V^2$, where $m_V$ is the mass of the vector meson. 
  In the RL truncation, the lowest masses are that of the $\rho$ and $\phi$ vector mesons.

\subsection{Numerical results for $G_h(Q^2)$ ($h=\eta,\,\eta'$)}

\begin{figure}[t]
\centering
\includegraphics[scale=0.325,keepaspectratio=true]{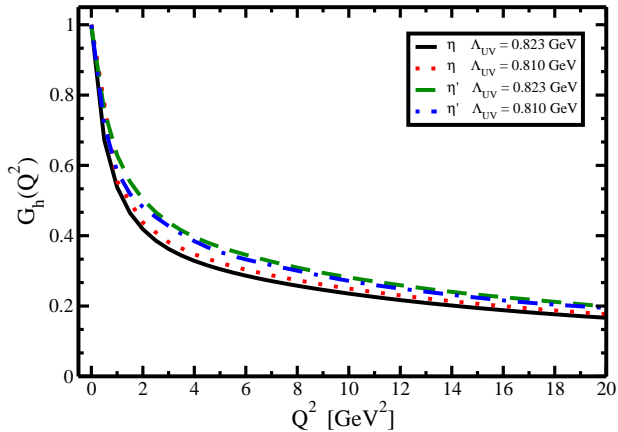}
\caption{\label{fig:TFF_1} Contact interactoin results for  transition form factors $G_\eta$ and
 $G_{\eta'}$, normalised to one at $Q^2=0$, for two values of the ultraviolet cutoff in the divergent
  integrals of the non-Abelian part of the kernel. Solid and dashed curves (color: black and red) correspond to $\eta$ and $\eta'$ transition form factors with $\Lambda_{UV}=0.823$ GeV,
  respectively, and dotted and dot-dash curves (color: green and blue) to the transition form factor with $\Lambda_{UV}\rightarrow \widetilde{\Lambda}_{UV}=0.810$ GeV, respectively.} 
\end{figure}

In \fig{fig:TFF_1} we present our contact interaction results for the $\gamma^*\gamma 
\to \eta$ and $\gamma^*\gamma  \to \eta'$  transition form factors, $G_\eta(Q^2)$ and 
$G_{\eta'}(Q^2)$, respectively, as a function of the photon virtuallity $Q^2$, up to 20 GeV$^2$.
In each case, we preset two curves, one where we use the same cutoff $\Lambda_{\text{UV}}$ as
in the RL contribution to the kernel and one where we modify it to be
$\widetilde{\Lambda}_{\text{UV}}$; see \tab{tab:K_A_Pars1}.
In the case of the static properties, changing $\Lambda_{\text{UV}}$ in the non-Abelian part
of the Kernel, gives a better description of masses and decay constants. However, as can be seen,
in the case of the transition form factor these two values give nearly equal results for both mesons.

\begin{figure}[t]
\centering
\includegraphics[scale=0.325,keepaspectratio=true]{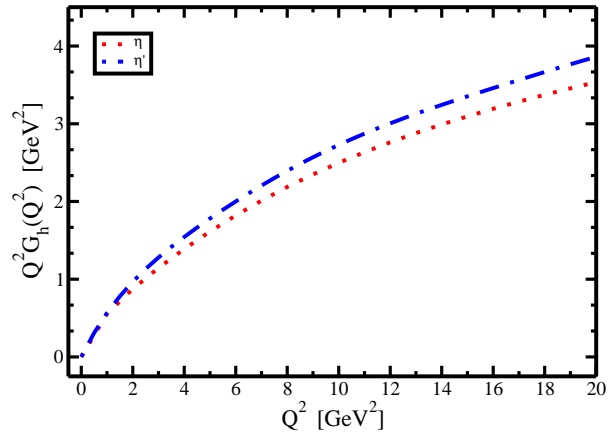}
\caption{\label{fig:TFFQ2_1} Contact interactoin results for $Q^2 G_\eta$ and
 $Q^2 G_{\eta'}$, for $\Lambda_{UV}=0.810$ GeV in in the divergent integrals of the 
 non-Abelian part of the kernel.} 
\end{figure}

From \fig{fig:TFF_1} we can see that both $G_\eta(Q^2)$ and $G_{\eta'}(Q^2)$ decrease as
functions of $Q^2$, decreasing rapidly for small values of $Q^2$. However, for larger values of
$Q^2$ ($> 4$ GeV$^2$),  this decreasing is slower for both $G_\eta(Q^2)$ and $G_{\eta'}(Q^2)$.
In \fig{fig:TFFQ2_1}, we plot  $Q^2 G_\eta$ and  $Q^2 G_{\eta'}$, for $\Lambda_{UV}=0.810$ GeV
in the divergent integrals of the non-Abelian part of the kernel. From Figs.~\ref{fig:TFF_1}
and ~\ref{fig:TFFQ2_1} we can see that both $G_\eta(Q^2)$ and $G_{\eta'}(Q^2)$ decrease 
slower than $Q^2$, in disagreement with perturbative QCD, which predicts that for very large 
$Q^2$ the product $Q^2 G_h(Q^2)$ becomes a constant~\cite{Lepage:1980fj,Lepage:1979zb,Efremov:1979qk}. This is not surprising for a contact interaction model, as we explain
below.
 \begin{figure}[t]
\centering
\includegraphics[scale=0.325,keepaspectratio=true]{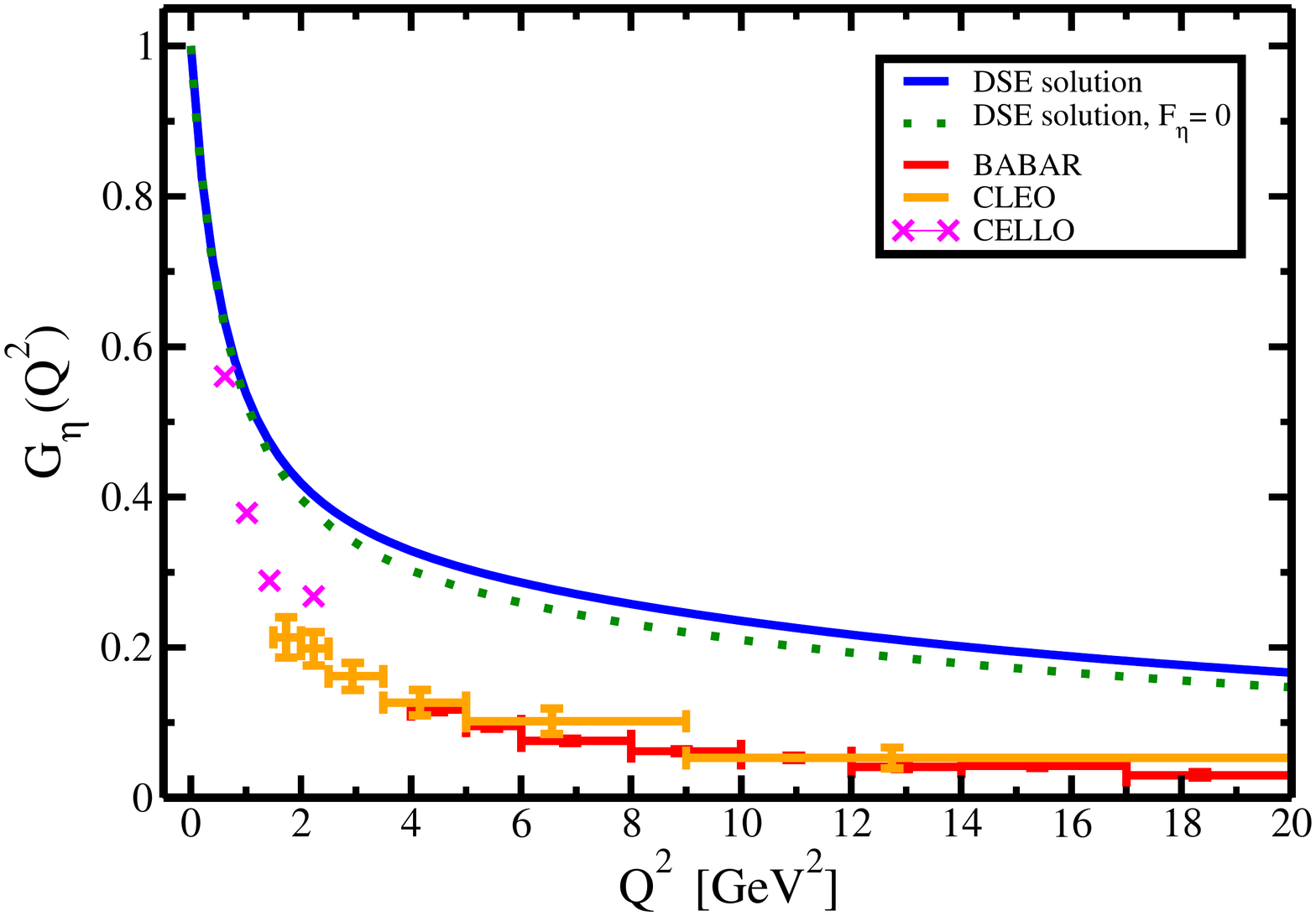}
\includegraphics[scale=0.325,keepaspectratio=true]{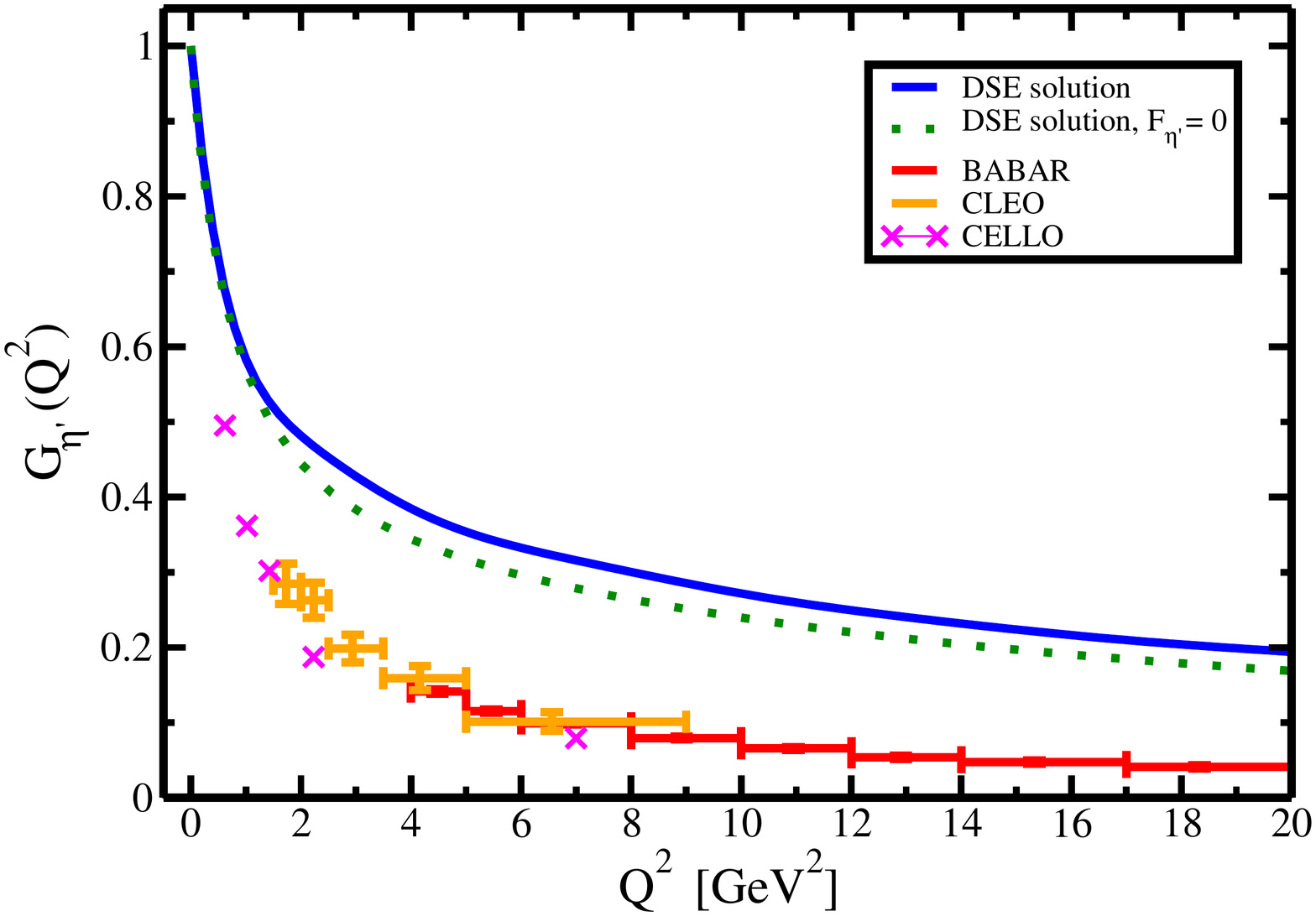}
\caption{\label{fig:TFF_2} Contact interactoin results for $G_\eta$  (top panel) and
 $Q^2 G_{\eta'}$ (bottom panel), for $\Lambda_{UV}=0.810$ GeV, compared to experimental 
 results from BABAR, CLEO, and CELLO.} 
\end{figure}
Finally, in \fig{fig:TFF_2} we compare our contact interaction results for $G_\eta$  (top panel)
 and $G_{\eta'}$ (bottom panel), with $\Lambda_{UV}=0.810$ GeV, to the experimental 
 results from CELLO~\cite{CELLO:1990klc}, CLEO~\cite{CLEO:1997fho}, and 
 BABAR~\cite{BaBar:2011nrp} collaborations. We note that the data has been normalized to one
 at $Q^2=0$. For this we need the transition form factors at $Q^2=0$. This can be extracted
  from the experimental values for the decay widths, using 
  $\Gamma_{h\to\gamma\gamma}= (1/4)\pi\alpha_{\text{em}}m_h^3|G_h(Q^2=0)|^2$ 
  to obtain $G_{\eta}(0)=0.2736$ and $G_{\eta'}(0)=0.3412$~\cite{Workman:2022ynf}
 
  Clearly, whilst capable of describing $\eta$ and $\eta'$ static properties, a contact
  interaction framework markedly disagrees with experimental data for $Q^2$ greater than 2 GeV$^2$,
  and produces transition form factors in conflict with perturbative QCD.
  This is not surprising since the contact interaction produces a momentum independent
   dressed-quark mass function~\cite{GutierrezGuerrero:2010md, Roberts:2010rn, Roberts:2011wy, 
 Roberts:2011cf, Chen:2012qr, Chen:2012txa, Segovia:2013uga, Segovia:2014aza, Xu:2015kta,
Bedolla:2015mpa, Bedolla:2016yxq, Raya:2017ggu, Serna:2016kdb, Serna:2017nlr,
Lu:2017cln, Gutierrez-Guerrero:2019uwa, Zhang:2020ecj, Gutierrez-Guerrero:2021rsx,
Xing:2021dwe, Cheng:2022jxe, Wang:2022mrh, Hernandez-Pinto:2023yin, Xing:2022jtt,
Xing:2023eed, Xing:2022mvk}  and also  Bethe-Salpeter equation kernels
   that are independent of the relative momentum between the quark and the 
   antiquark~\cite{GutierrezGuerrero:2010md, Roberts:2010rn, Roberts:2011wy, 
 Roberts:2011cf, Chen:2012qr, Chen:2012txa, Segovia:2013uga, Segovia:2014aza, Xu:2015kta,
Bedolla:2015mpa, Bedolla:2016yxq, Raya:2017ggu, Serna:2016kdb, Serna:2017nlr,
Lu:2017cln, Gutierrez-Guerrero:2019uwa, Zhang:2020ecj, Gutierrez-Guerrero:2021rsx,
Xing:2021dwe, Cheng:2022jxe, Wang:2022mrh, Hernandez-Pinto:2023yin, Xing:2022jtt,
Xing:2023eed, Xing:2022mvk}, in contrast to QCD-based SDE studies and lattice QCD,
   These two features are fundamentally the source of the discrepancy between the elastic 
   and transition form factor obtained with a  contact interaction and those provided by experiment. 
   However, our work highlights, and complements others~\cite{Gutierrez-Guerrero:2019uwa,
    Roberts:2010rn, Roberts:2011wy,Bedolla:2015mpa,Bedolla:2016yxq,Wang:2022mrh}, 
   that eleastic and transition form factor observables, are very sensitive to the
   running (with momentum) of the dressed-quark mass.

\section{\label{sec:summary}Summary and conclusions}

We have constructed a contact interaction model for the $\eta$ and $\eta'$ mesons in a 
SDE-BSE approach to QCD and computed masses, decay widths and transition form factors.
Tables~\ref{tab:eta-etaprime_results} and ~\ref{tab:eta-etaprime_decay_const}  show that
this model gives an excellent description of the $\eta$ and $\eta'$ static properties,
namely their masses, decay width and decay constants. 
Although contact interaction results for the $\eta$ and $\eta'$ transition form factors markedly
 disagree with experimental data for $Q^2$ greater than 2 GeV$^2$, and produces transition form
  factors  in conflict with perturbative QCD, the fact that this model gives a very good description of
 the $\eta$ and $\eta'$ static properties makes it an excellent tool for the investigation of $U_A(1)$
  symmetry restoration at finite temperature and density in the SDE-BSE approach to QCD. 
  Work on this direction is under way and will be reported elsewhere.

\section*{Acknowledgements}

JJCM acknowledges financial support fromthe University of Sonora under grant
USO315007861. ECM and BAZ acknowledge financial support from CONACyT  for postgraduate
studies. JS acknowledges financial support from Ministerio Espa\~nol de Ciencia e 
Innovaci\'on under grant No. PID2019-107844GB-C22 and from the Junta de Andaluc\'ia 
under contract Nos. Operativo FEDER Andaluc\'ia 2014-2020 UHU-1264517, P18-FR-5057 
and also PAIDI FQM-370.


\begin{thebibliography}{100}


\bibitem{Roberts:2022rxm}
C.~D.~Roberts,
EPJ Web Conf. \textbf{282}, 01006 (2023)
[arXiv:2211.09905 [hep-ph]].


\bibitem{Roberts:2023lap}
C.~D.~Roberts,
[arXiv:2304.00154 [hep-ph]].


\bibitem{Roberts:1994dr}
C.~D.~Roberts and A.~G.~Williams,
Prog. Part. Nucl. Phys. \textbf{33}, 477-575 (1994)
[arXiv:hep-ph/9403224 [hep-ph]].

\bibitem{Maris:2003vk}
P.~Maris and C.~D.~Roberts,
Int. J. Mod. Phys. E \textbf{12}, 297-365 (2003)
[arXiv:nucl-th/0301049 [nucl-th]].


\bibitem{Bashir:2012fs}
A.~Bashir, L.~Chang, I.~C.~Cloet, B.~El-Bennich, Y.~X.~Liu, C.~D.~Roberts and P.~C.~Tandy,
Commun. Theor. Phys. \textbf{58} (2012), 79-134
[arXiv:1201.3366 [nucl-th]].


\bibitem{Fischer:2018sdj}
C.~S.~Fischer,
Prog. Part. Nucl. Phys. \textbf{105}, 1-60 (2019)
doi:10.1016/j.ppnp.2019.01.002
[arXiv:1810.12938 [hep-ph]].


\bibitem{GutierrezGuerrero:2010md}
  L.~X.~Gutierrez-Guerrero, A.~Bashir, I.~C.~Cloet and C.~D.~Roberts,
  Phys.\ Rev.\ C {\bf 81}, 065202 (2010)
  [arXiv:1002.1968 [nucl-th]].


\bibitem{Roberts:2010rn}
  H.~L.~L.~Roberts, C.~D.~Roberts, A.~Bashir, L.~X.~Gutierrez-Guerrero and P.~C.~Tandy,
  Phys.\ Rev.\ C {\bf 82}, 065202 (2010)
  [arXiv:1009.0067 [nucl-th]].
   
  
\bibitem{Roberts:2011wy}
  H.~L.~L.~Roberts, A.~Bashir, L.~X.~Gutierrez-Guerrero, C.~D.~Roberts and D.~J.~Wilson,
  Phys.\ Rev.\ C {\bf 83}, 065206 (2011)
  [arXiv:1102.4376 [nucl-th]].

\bibitem{Roberts:2011cf}
  H.~L.~L.~Roberts, L.~Chang, I.~C.~Cloet and C.~D.~Roberts,
  Few Body Syst.\  {\bf 51}, 1 (2011)
  [arXiv:1101.4244 [nucl-th]].


\bibitem{Chen:2012qr}
  C.~Chen, L.~Chang, C.~D.~Roberts, S.~Wan and D.~J.~Wilson,
  Few Body Syst.\  {\bf 53}, 293 (2012)
  [arXiv:1204.2553 [nucl-th]].

\bibitem{Chen:2012txa}
C.~Chen, L.~Chang, C.~D.~Roberts, S.~M.~Schmidt, S.~Wan and D.~J.~Wilson,
Phys. Rev. C \textbf{87}, 045207 (2013)
[arXiv:1212.2212 [nucl-th]].



\bibitem{Segovia:2013uga}
J.~Segovia, C.~Chen, I.~C.~Clo\"et, C.~D.~Roberts, S.~M.~Schmidt and S.~Wan,
Few Body Syst. \textbf{55}, 1-33 (2014)
doi:10.1007/s00601-013-0734-x
[arXiv:1308.5225 [nucl-th]].



\bibitem{Segovia:2014aza}
J.~Segovia, I.~C.~Cloet, C.~D.~Roberts and S.~M.~Schmidt,
Few Body Syst. \textbf{55}, 1185-1222 (2014)
doi:10.1007/s00601-014-0907-2
[arXiv:1408.2919 [nucl-th]].



\bibitem{Xu:2015kta}
S.~S.~Xu, C.~Chen, I.~C.~Cloet, C.~D.~Roberts, J.~Segovia and H.~S.~Zong,
Phys. Rev. D \textbf{92}, no.11, 114034 (2015)
doi:10.1103/PhysRevD.92.114034
[arXiv:1509.03311 [nucl-th]].


\bibitem{Bedolla:2015mpa}
M.~A.~Bedolla, J.~J.~Cobos-Mart\'\i{}nez and A.~Bashir,
Phys. Rev. D \textbf{92}, no.5, 054031 (2015)
doi:10.1103/PhysRevD.92.054031
[arXiv:1601.05639 [hep-ph]].


\bibitem{Bedolla:2016yxq}
  M.~A.~Bedolla, K.~Raya, J.~J.~Cobos-Mart\'inez and A.~Bashir,
  Phys.\ Rev.\ D {\bf 93}, no. 9, 094025 (2016)
  [arXiv:1606.03760 [hep-ph]].
  
  
\bibitem{Raya:2017ggu}
  K.~Raya, M.~A.~Bedolla, J.~J.~Cobos-Mart\'inez and A.~Bashir,
  Few Body Syst.\  {\bf 59}, no. 6, 133 (2018)
  [arXiv:1711.00383 [nucl-th]].

 
\bibitem{Serna:2016kdb}
F.~E.~Serna, M.~A.~Brito and G.~Krein,
AIP Conf. Proc. \textbf{1701}, no.1, 100018 (2016)
doi:10.1063/1.4938727
[arXiv:1607.03823 [nucl-th]].



\bibitem{Serna:2017nlr}
F.~E.~Serna, B.~El-Bennich and G.~Krein,
Phys. Rev. D \textbf{96}, no.1, 014013 (2017)
doi:10.1103/PhysRevD.96.014013
[arXiv:1703.09181 [hep-ph]].

\bibitem{Lu:2017cln}
Y.~Lu, C.~Chen, C.~D.~Roberts, J.~Segovia, S.~S.~Xu and H.~S.~Zong,
Phys. Rev. C \textbf{96}, no.1, 015208 (2017)
doi:10.1103/PhysRevC.96.015208
[arXiv:1705.03988 [nucl-th]].


\bibitem{Gutierrez-Guerrero:2019uwa}
L.~X.~Guti\'errez-Guerrero, A.~Bashir, M.~A.~Bedolla and E.~Santopinto,
Phys. Rev. D \textbf{100}, no.11, 114032 (2019)
[arXiv:1911.09213 [nucl-th]].

\bibitem{Zhang:2020ecj}
J.~L.~Zhang, Z.~F.~Cui, J.~Ping and C.~D.~Roberts,
Eur. Phys. J. C \textbf{81}, no.1, 6 (2021)
doi:10.1140/epjc/s10052-020-08791-1
[arXiv:2009.11384 [hep-ph]].


\bibitem{Gutierrez-Guerrero:2021rsx}
L.~X.~Guti\'errez-Guerrero, G.~Paredes-Torres and A.~Bashir,
Phys. Rev. D \textbf{104}, no.9, 094013 (2021)
doi:10.1103/PhysRevD.104.094013
[arXiv:2109.09058 [hep-ph]].

\bibitem{Xing:2021dwe}
Z.~Xing, K.~Raya and L.~Chang,
Phys. Rev. D \textbf{104}, no.5, 054038 (2021)
doi:10.1103/PhysRevD.104.054038
[arXiv:2107.05158 [nucl-th]].

\bibitem{Cheng:2022jxe}
P.~Cheng, F.~E.~Serna, Z.~Q.~Yao, C.~Chen, Z.~F.~Cui and C.~D.~Roberts,
Phys. Rev. D \textbf{106}, no.5, 054031 (2022)
doi:10.1103/PhysRevD.106.054031
[arXiv:2207.13811 [hep-ph]].


\bibitem{Wang:2022mrh}
X.~Wang, Z.~Xing, J.~Kang, K.~Raya and L.~Chang,
Phys. Rev. D \textbf{106}, no.5, 054016 (2022)
[arXiv:2207.04339 [hep-ph]].

\bibitem{Hernandez-Pinto:2023yin}
R.~J.~Hern\'andez-Pinto, L.~X.~Guti\'errez-Guerrero, A.~Bashir, M.~A.~Bedolla and I.~M.~Higuera-Angulo,
Phys. Rev. D \textbf{107}, no.5, 054002 (2023)
doi:10.1103/PhysRevD.107.054002
[arXiv:2301.11881 [hep-ph]].

\bibitem{Xing:2022jtt}
Z.~Xing and L.~Chang,
Phys. Rev. D \textbf{107}, no.1, 014019 (2023)
doi:10.1103/PhysRevD.107.014019
[arXiv:2210.12452 [hep-ph]].

\bibitem{Xing:2023eed}
Z.~Xing, M.~Ding, K.~Raya and L.~Chang,
[arXiv:2301.02958 [hep-ph]].


\bibitem{Xing:2022mvk}
Z.~Xing, M.~Ding and L.~Chang,
Phys. Rev. D \textbf{107}, no.3, L031502 (2023)
doi:10.1103/PhysRevD.107.L031502
[arXiv:2211.06635 [hep-ph]].


\bibitem{Ahmad:2020jzn}
A.~Ahmad, A.~Bashir, M.~A.~Bedolla and J.~J.~Cobos-Mart\'\i{}nez,
J. Phys. G \textbf{48}, no.7, 075002 (2021)
doi:10.1088/1361-6471/abd88f
[

\bibitem{Serna:2016ifh}
F.~E.~Serna and G.~Krein,
EPJ Web Conf. \textbf{137}, 13015 (2017)
doi:10.1051/epjconf/201713713015
[arXiv:1612.00473 [nucl-th]].

\bibitem{Wang:2013wk}
K.~l.~Wang, Y.~x.~Liu, L.~Chang, C.~D.~Roberts and S.~M.~Schmidt,
Phys. Rev. D \textbf{87}, no.7, 074038 (2013)
doi:10.1103/PhysRevD.87.074038
[arXiv:1301.6762 [nucl-th]].


\bibitem{Bowman:2004jm}
P.~O.~Bowman, U.~M.~Heller, D.~B.~Leinweber, M.~B.~Parappilly and A.~G.~Williams,
Phys. Rev. D \textbf{70}, 034509 (2004)
doi:10.1103/PhysRevD.70.034509
[arXiv:hep-lat/0402032 [hep-lat]].

\bibitem{Dudal:2008sp}
D.~Dudal, J.~A.~Gracey, S.~P.~Sorella, N.~Vandersickel and H.~Verschelde,
Phys. Rev. D \textbf{78}, 065047 (2008)
doi:10.1103/PhysRevD.78.065047
[arXiv:0806.4348 [hep-th]].

\bibitem{Huber:2010cq}
M.~Q.~Huber, R.~Alkofer and S.~P.~Sorella,
AIP Conf. Proc. \textbf{1343}, 158-160 (2011)
doi:10.1063/1.3574962
[arXiv:1010.4802 [hep-th]].

\bibitem{Boucaud:2011ug}
P.~Boucaud, J.~P.~Leroy, A.~L.~Yaouanc, J.~Micheli, O.~Pene and J.~Rodriguez-Quintero,
Few Body Syst. \textbf{53}, 387-436 (2012)
doi:10.1007/s00601-011-0301-2
[arXiv:1109.1936 [hep-ph]].

\bibitem{Ayala:2012pb}
A.~Ayala, A.~Bashir, D.~Binosi, M.~Cristoforetti and J.~Rodriguez-Quintero,
Phys. Rev. D \textbf{86}, 074512 (2012)
doi:10.1103/PhysRevD.86.074512
[arXiv:1208.0795 [hep-ph]].

\bibitem{Gao:2017uox}
F.~Gao, S.~X.~Qin, C.~D.~Roberts and J.~Rodriguez-Quintero,
Phys. Rev. D \textbf{97}, no.3, 034010 (2018)
doi:10.1103/PhysRevD.97.034010
[arXiv:1706.04681 [hep-ph]].

\bibitem{Ebert:1996vx}
  D.~Ebert, T.~Feldmann and H.~Reinhardt,
  Phys.\ Lett.\ B {\bf 388}, 154 (1996)
  [hep-ph/9608223].

\bibitem{Nambu:1950dpa}
Y.~Nambu,
Prog. Theor. Phys. \textbf{5}, 614-633 (1950)

\bibitem{Salpeter:1951sz}
E.~E.~Salpeter and H.~A.~Bethe,
Phys. Rev. \textbf{84}, 1232-1242 (1951)

\bibitem{Gell-Mann:1951ooy}
M.~Gell-Mann and F.~Low,
Phys. Rev. \textbf{84}, 350-354 (1951)

\bibitem{Nakanishi:1969ph}
N.~Nakanishi,
Prog. Theor. Phys. Suppl. \textbf{43}, 1-81 (1969)
doi:10.1143/PTPS.43.1


\bibitem{Maris:1997hd}
P.~Maris, C.~D.~Roberts and P.~C.~Tandy,
Phys. Lett. B \textbf{420}, 267-273 (1998)
[arXiv:nucl-th/9707003 [nucl-th]].

\bibitem{Hutauruk:2019ipp}
P.~T.~P.~Hutauruk, J.~J.~Cobos-Mart\'\i{}nez, Y.~Oh and K.~Tsushima,
Phys. Rev. D \textbf{100}, no.9, 094011 (2019)
doi:10.1103/PhysRevD.100.094011
[arXiv:1908.02406 [hep-ph]].

\bibitem{Workman:2022ynf}
R.~L.~Workman \textit{et al.} [Particle Data Group],
PTEP \textbf{2022}, 083C01 (2022)


\bibitem{Ding:2018xwy}
M.~Ding, K.~Raya, A.~Bashir, D.~Binosi, L.~Chang, M.~Chen and C.~D.~Roberts,
Phys. Rev. D \textbf{99}, no.1, 014014 (2019)
[arXiv:1810.12313 [nucl-th]].

\bibitem{Feldmann:1998sh}
T.~Feldmann, P.~Kroll and B.~Stech,
Phys. Lett. B \textbf{449}, 339-346 (1999)
doi:10.1016/S0370-2693(99)00085-4
[arXiv:hep-ph/9812269 [hep-ph]].

\bibitem{Feldmann:1998vh}
T.~Feldmann, P.~Kroll and B.~Stech,
Phys. Rev. D \textbf{58}, 114006 (1998)
doi:10.1103/PhysRevD.58.114006
[arXiv:hep-ph/9802409 [hep-ph]].

\bibitem{Feldmann:1999uf}
T.~Feldmann,
Int. J. Mod. Phys. A \textbf{15}, 159-207 (2000)
doi:10.1142/S0217751X00000082
[arXiv:hep-ph/9907491 [hep-ph]].

\bibitem{Nakanishi:1965zza}
N.~Nakanishi,
Phys. Rev. \textbf{138}, B1182-B1192 (1965)
doi:10.1103/PhysRev.138.B1182

\bibitem{Nakanishi:1965zz}
N.~Nakanishi,
Phys. Rev. \textbf{139}, B1401-B1406 (1965)
doi:10.1103/PhysRev.139.B1401

\bibitem{Osipov:2014dya}
A.~A.~Osipov, B.~Hiller and A.~H.~Blin,
Acta Phys. Polon. Supp. \textbf{8}, no.1, 183 (2015)
[arXiv:1411.2137 [hep-ph]].

\bibitem{Takizawa:1995ku}
M.~Takizawa, Y.~Nemoto and M.~Oka,
Austral. J. Phys. \textbf{50} (1997), 187-197
[arXiv:hep-ph/9602346 [hep-ph]].


\bibitem{Lepage:1980fj}
G.~P.~Lepage and S.~J.~Brodsky,
Phys. Rev. D \textbf{22}, 2157 (1980)

\bibitem{Lepage:1979zb}
G.~P.~Lepage and S.~J.~Brodsky,
Phys. Lett. B \textbf{87}, 359-365 (1979)

\bibitem{Efremov:1979qk}
A.~V.~Efremov and A.~V.~Radyushkin,
Phys. Lett. B \textbf{94}, 245-250 (1980)


\bibitem{CELLO:1990klc}
H.~J.~Behrend \textit{et al.} [CELLO],
Z. Phys. C \textbf{49}, 401-410 (1991)

\bibitem{CLEO:1997fho}
J.~Gronberg \textit{et al.} [CLEO],
Phys. Rev. D \textbf{57}, 33-54 (1998)
[arXiv:hep-ex/9707031 [hep-ex]].

\bibitem{BaBar:2011nrp}
P.~del Amo Sanchez \textit{et al.} [BaBar],
Phys. Rev. D \textbf{84}, 052001 (2011)
[arXiv:1101.1142 [hep-ex]].




 

 \end{thebibliography}
\end{document}